\documentclass[]{spie}  

\usepackage{amsmath,amsfonts,amssymb}
\usepackage{graphicx}
\usepackage[colorlinks=true, allcolors=blue]{hyperref}

\title{Optical throughput and sensitivity of JWST NIRSpec}

\author[a]{Giovanna Giardino}
\author[b]{Rachana Bhatawdekar}
\author[c]{Stephan M. Birkmann}
\author[d]{Pierre Ferruit}
\author[c]{Timothy Rawle}
\author[d]{Catarina Alves de Oliveira}
\author[c]{Torsten B\"{o}ker}
\author[h]{Peter Jakobsen}
\author[e]{Nimisha Kumari}
\author[f]{Marcos L\'{o}pez-Caniego}
\author[c]{Nora L\"{u}tzgendorf}
\author[e]{Elena Manjavacas}
\author[g]{Charles Proffitt}
\author[c]{Marco Sirianni}
\author[c]{Maurice Te Plate}
\author[e]{Peter Zeidler}
\affil[a]{ATG Europe for the European Space Agency, ESTEC, The Netherlands}
\affil[b]{European Space Agency, ESTEC, Noordwijk, The Netherlands}
\affil[c]{European Space Agency, STScI, Baltimore, USA}
\affil[d]{European Space Agency, ESAC, Madrid, Spain}
\affil[e]{AURA for the European Space Agency, STScI, Baltimore, USA}
\affil[f]{Aurora Technology for the European Space Agency, ESAC, Madrid, Spain}
\affil[g]{Space Telescope Science Institute, Baltimore, USA}
\affil[h]{Cosmic Dawn Center, Niels Bohr Institute, Copenhagen, Denmark}
\authorinfo{Further author information: (Send correspondence to G.G.)\\G.G.: E-mail: Giovanna.Giardino@esa.int, Telephone: +31 71 565 4625}

\begin{document} 
\maketitle

\begin{abstract}
  
To achieve its ambitious scientific goals, the Near-Infrared
Spectrograph, NIRSpec, on board the Webb Space Telescope, needs to
meet very demanding throughput requirements, here quantified in terms
of photon-conversion efficiency (PCE). During the calibration
activities performed for the instrument commissioning, we have obtained
the first in-flight measurements of its PCE and also updated the
modeling of the light losses occurring in the NIRSpec slit devices.

The measured PCE of NIRSpec fixed-slit and multi-object spectroscopy
modes overall meets or exceeds the pre-launch model predictions. The
results are more contrasted for the integral-field spectroscopy mode,
where the differences with the model can reach $-$20\%, above 4~$\mu$m,
and exceed $+$30\%, below 2~$\mu$m. Additionally, thanks to the
high quality of the JWST point-spread function, our slit-losses, at
the shorter wavelength, are significantly decreased with respect to
the pre-flight modeling.

These results, combined with the confirmed
low noise performance of the detectors, make of NIRSpec an
exceptionally sensitive spectrograph.

\end{abstract}


\section{INTRODUCTION}
\label{sec:intro}  

The Near-Infrared Spectrograph (NIRSpec) is one the four focal plane
instruments on the Webb Space Telescope; its
main scientific goal is to enable multi-object spectroscopy of
faint high redshift galaxies and thus advance our understanding of the
early stage of galaxy formation at the dawn of our universe.
Nevertheless its innovative design allows sensitive observations of a
wide range of targets to be performed, including detailed
spectroscopic studies of the atmospheres of exoplanets, to choose an
example at the opposite end of
the Universe distance scale[\citenum{pfa+2022}].

To support these studies, NIRSpec is capable of carrying out
low, medium and high resolution spectroscopy,
both in single- object mode employing any one of five fixed slits
(FS), or a 3.1$\times$3.2 arcsec$^2$ integral field unit (IFU) for
Integral Field Spectroscopy observations (IFS), or in multi-object
(MOS) mode employing a novel programmable micro-shutter array (MSA)
covering a 3.6$\times$3.4 arcmin$^2$ field of
view[\citenum{pfa+2022}].  NIRSpec is equipped with a double
pass-prism for low-resolution spectroscopy over the entire
wavelength-range, 0.6 $-$ 5.3 $\mu$m, and two sets of three
diffraction gratings providing spectral resolutions of $R{\simeq}1000$
and $R{\simeq}2700$, respectively. The two sets of gratings are
formally specified to cover the 1.0-5.0\!~$\mu$m wavelength region in
each of three overlapping bands: 1.0-1.8\!~$\mu$m (Band~I),
1.7-3.0\!~$\mu$m (Band~II) and 2.9-5.0\!~$\mu$m (Band~III). The
nominal instrument configurations, and corresponding combination of
dispersive elements and filters, are listed in
\autoref{tab:NIRS_modes}.  The shorter wavelength F070LP long-pass
filter listed is included to enable observations below 1\!~$\mu$m with
the G140M and G140H gratings, which are free of contaminating
second-order light only up to a wavelength of 1.27\!~$\mu$m.

\begin{table}[ht]
	\caption{Nominal NIRSpec science configurations}
  	\label{tab:NIRS_modes}
        \begin{center}
	\begin{tabular}{|c|c|c|c|c|}
	\hline\hline
	Band & Disperser element & Resolution $\lambda / \Delta\lambda$ & Filter & Spectral range / $\mu$m\\
	\hline
        \rule[-1ex]{0pt}{3.5ex} 0 & G140M, G140H & 1000, 2700 & F070LP & 0.7--1.2 \\
        \hline
	\rule[-1ex]{0pt}{3.5ex} I & G140M, G140H & 1000, 2700 & F100LP & 1.0--1.8 \\
        \hline
	\rule[-1ex]{0pt}{3.5ex} II & G235M, G235H & 1000, 2700 & F170LP & 1.7--3.1 \\
        \hline
	\rule[-1ex]{0pt}{3.5ex} III & G395M, G395H & 1000, 2700 & F290LP & 2.9--5.2 \\
        \hline
	\rule[-1ex]{0pt}{3.5ex} n/a & PRISM & ~100 & CLEAR & 0.6--5.3 \\
	\hline
	\end{tabular}
        \end{center}
\end{table}

 \subsection{Photon-conversion efficiency}
 


A key performance parameter in every optical instrument, either a
telescope or microscope, is its efficiency: how many photons will make
it through the optical chain the instrument is comprised of. Barring
systematics, the throughput and the noise performance of the detectors
are the parameters driving the photometric sensitivity of an
instrument.

NIRSpec is a complex instrument that supports many different modes of
observations. Its optical train is reflective throughout, save for the
order-separation filters and the low resolution dispersive prism. The
three primary optical modules of NIRSpec are each implemented in the
form of three-mirror anastigmats employing high-order aspherical
surfaces[\citenum{geyl11}]. Counting the five plane fold mirrors and
the disperser, the light entering NIRSpec undergoes a total of 15
reflections (in the FS/MOS mode) before reaching the detector array.
Coupled with the telescope element of JWST, the total number of
reflections a photon goes thorough in its path from the primary mirror
to the detectors at the NIRSpec focal plane is 19! The IFU mode adds a
further 8 gold-coated reflections to the light path. Ultimately,
though, we do not count photons but the electrons generated in the
semi-conductor substrate of the detectors, with their own performance
in terms of quantum efficiency (QE). For this reason, we describe the
throughput of our instrument in terms of Photon Conversion Efficiency,
PCE.

To achieve its ambitious science goals, NIRSpec was designed with
demanding requirements in terms of optical efficiency  across
its entire wavelength range - and therefore particular attention
was placed in the design and manufacturing of its reflective elements,
making sure that all the optical surfaces received the highest quality
optical coatings and were kept extremely clean during instrument
assembly.  With the aim of achieving high optical throughput at the
blue end, all mirrors are coated with protected silver.
Additionally, the two Teledyne H2RG focal plane detectors have excellent QE, overall
exceeding 70\% , as illustrated by Rauscher et al.\,[\citenum{bernie2014}].

Preliminary measurements of the NIRSpec PCE were obtained during the
NIRSpec ground-testing campaign that took place in 2013. The measurements,
though, were affected by significant uncertainties (due to 
limitations of the ground-testing equipment), and, in addition, in
2015 the detector arrays were replaced with new
devices. Therefore, before flight, the PCE performance of NIRSpec was
based on a radiometric model of the instrument obtained by combining
the measured absolute QE of the replaced arrays with
individual-component measurements made on each optical element or
appropriate witness samples, together with the nominal reflectivity of
the telescope optics[\citenum{light16}]. The PCE values predicted
before flight were very high, reaching in excess of 50\% in prism
configuration, and at peak blaze in all six gratings - see Jakobsen et
al.[\citenum{pfa+2022}].

The moment of truth finally came during the commissioning of the
instrument, with the first observations of a spectro-photometric
standard star.

\section{OBSERVATIONS AND PROCESSING}

JWST was launched on the 25th of December of 2021. After deployment
and reaching the target halo-orbit around the Sun-Earth L2 point, the
telescope and instrumentation underwent a six months period of
commissioning activities to prepare the observatory for its scientific
observations. During this phase, NIRSpec commissioning team
commanded NIRSpec to perform internal calibration exposures and
on-sky observations, acquiring the data necessary to assess the instrument's in-flight
performance and generate calibration reference data[\citenum{tspie+2022}].

For the absolute spectro-photometric calibration of the instrument, we
observed the standard stars 1808347 (TYC 4433-1800-1), an A3 star
selected from STScI Calibration database CALSPEC, as well as two other
reference stars during one of the first instrument check-out exposures
on sky: wd1057+719 (a white dwarf) and p177d (G0),
using all the dispersers, for a total of 9 filter-disperser
combinations\footnote{Proposal ID (PID) 1128}. To asses the PCE for
the MOS and FS mode, the star was observed through the square 1.6
arcsec-wide slit (S1600A1), which has minimal geometrical loss, with
typical values of 3--9\% (see Sect.\,\ref{sec:losses}).  To enable the
subtraction of the background emission a two-point 1-arcsec dither in the
spatial direction was executed within the slit, in each configuration,
while, in the case of the IFS mode, a 4-point nod was executed across the
IFU aperture field-of-view.

\begin{figure} [ht]
\begin{center}
\begin{tabular}{ll} 
  \includegraphics[height=4.5cm]{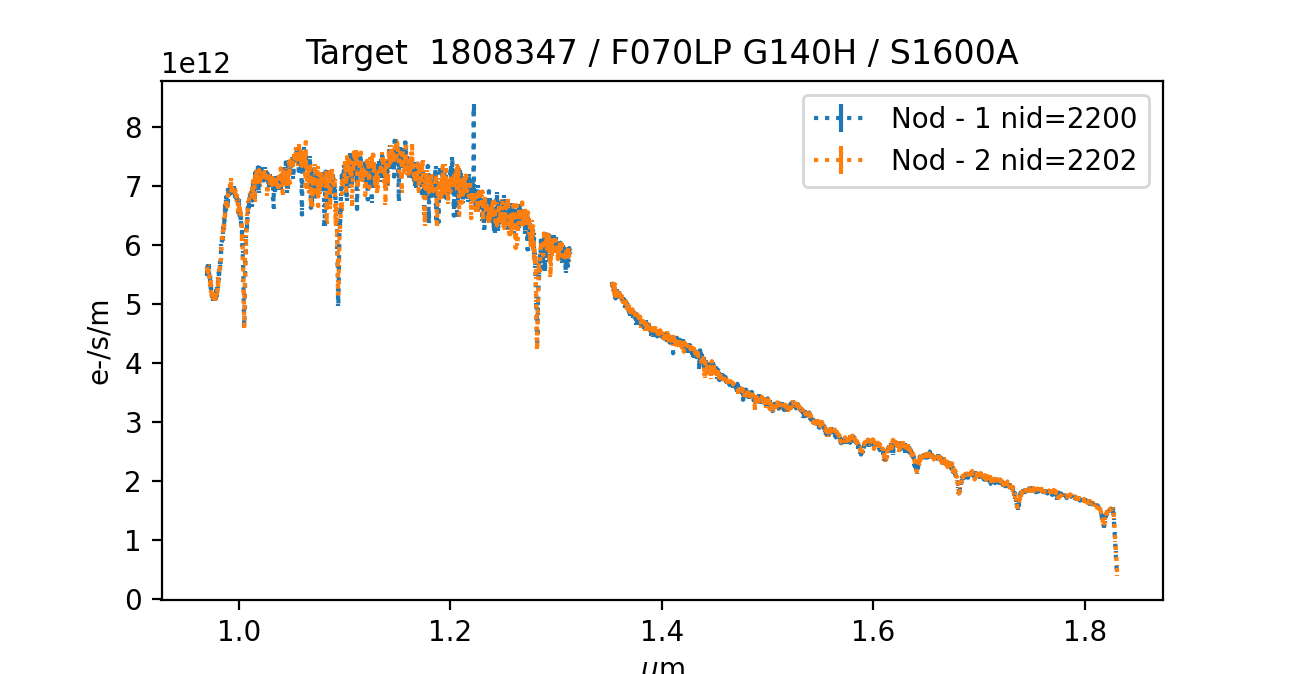}
  \includegraphics[height=4.5cm]{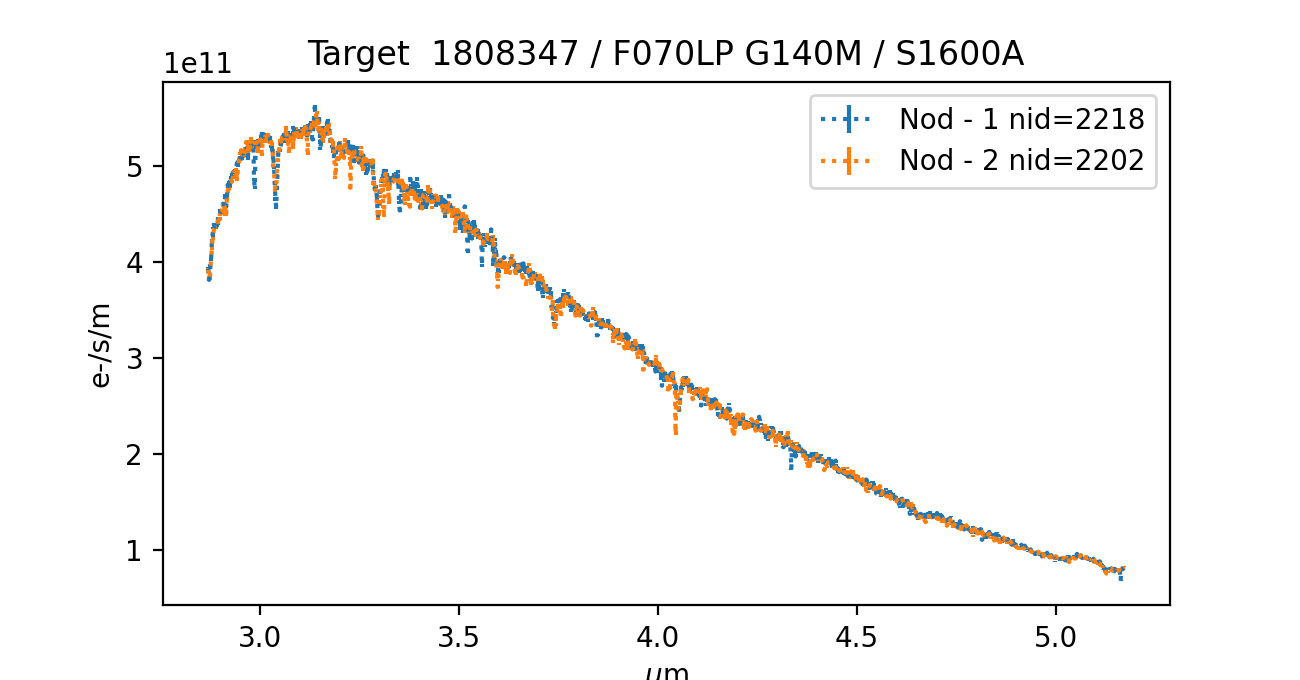}
\end{tabular}
\end{center}
\caption[example] { \label{fig:example} Example of extracted spectra
  in units of electron rate per unit wavelength for two of the
  disperser-filter configurations. The gap in the high-resolution
  spectrum (left-hand panel) is due to the gap between the two NIRSpec detectors. Minimal
  processing was applied to these data -- see text.}
\end{figure}

The exposures were processed with our NIRSpec ramp-to-slope
pipeline that performs the following basic data reduction steps: bias
subtraction, reference pixel subtraction, linearity correction, dark
subtraction and finally count-rate estimation, including jump
detection and cosmic-ray rejection -- see Birkmann et
al.[\citenum{st_spie+2022}] for more details on this last step. From
the count-rate images the wavelength calibrated spectra were obtained
using the Stage 2 of the NIRSpec pipeline to perform the following
operations for the FS: subtract background (combining the exposures
from the two-nods); extract sub-image containing the spectral trace
and assign wavelength and spatial coordinates to each pixel therein;
generate a rectified spectrum re-sampled on regular 2D-grid; compute
the 1D-spectrum obtained by spatial integration. Figure
\ref{fig:example} shows the spectra for configurations F070LP-G140H
and F290-G395M, as examples. For the IFS exposures, after background
subtractions and rectification of the 30 IFU traces, Stage 2 of NIRSpec
pipeline builds the three-dimensional data cube from where the
1D-spectrum is obtained by spatial integration of a circular area.

The PCE of the instrument through the given aperture is given simply by
dividing the observed number of electron rate per wavelength element by
the value of photons flux collected by the primary mirror (per wavelength element).

\section{RESULTS}

\begin{figure}[ht]
\begin{center}
\begin{tabular}{c} 
  \includegraphics[height=6.5cm]{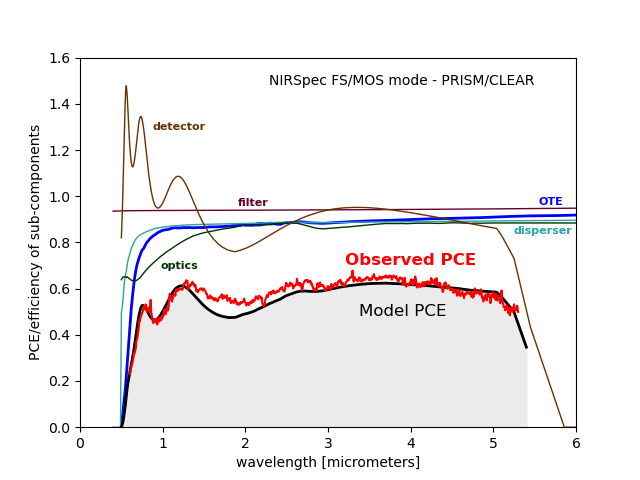}
  \includegraphics[height=6.5cm]{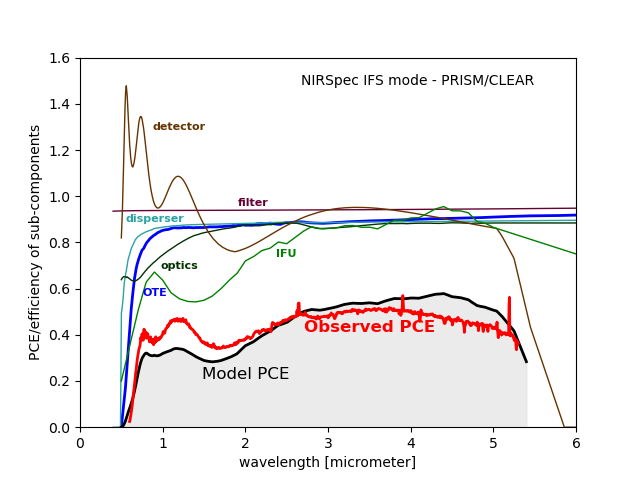}
\end{tabular}
\end{center}
\caption[pce_prism] { \label{fig:prismpce} The measured PCE as a
  function of wavelength of  the low resolution configurations for the
  FS/MOS mode on the left and the IFS mode on the right. The data are
  here compared with the predictions by our radiometric model of the
  instrument. The model includes the reflectivity of the telescope
  optics (marked OTE) and the responsive quantum efficiency of the
  NIRSpec detector array, but excludes the light losses occurring in
  the slit device employed. The measured values on the other-hand are
  affected by path-losses (for which we have not applied any
  correction).}
\end{figure}

\begin{figure}
\begin{center}
\begin{tabular}{cc} 
  \includegraphics[height=6cm]{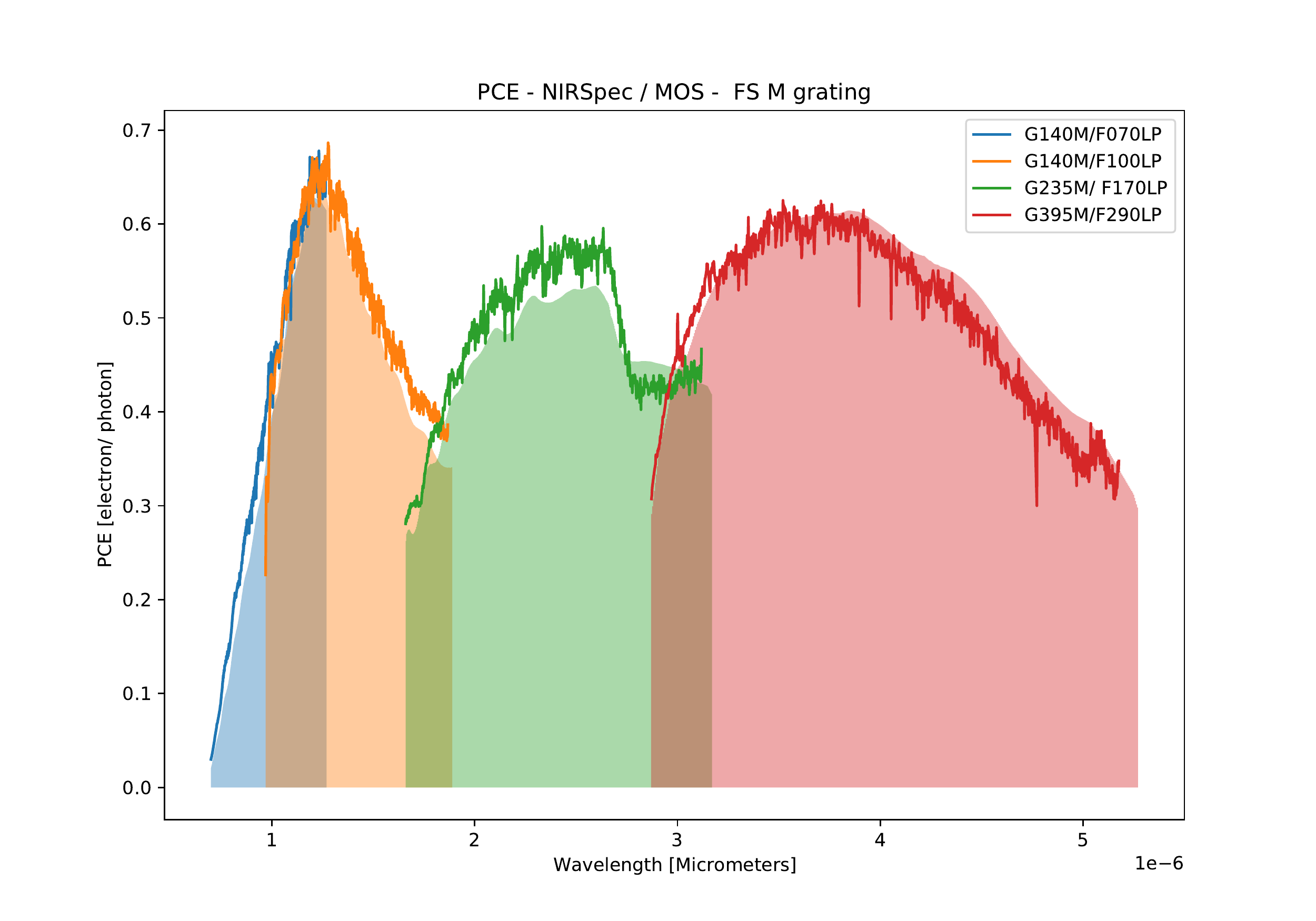}
  \includegraphics[height=6cm]{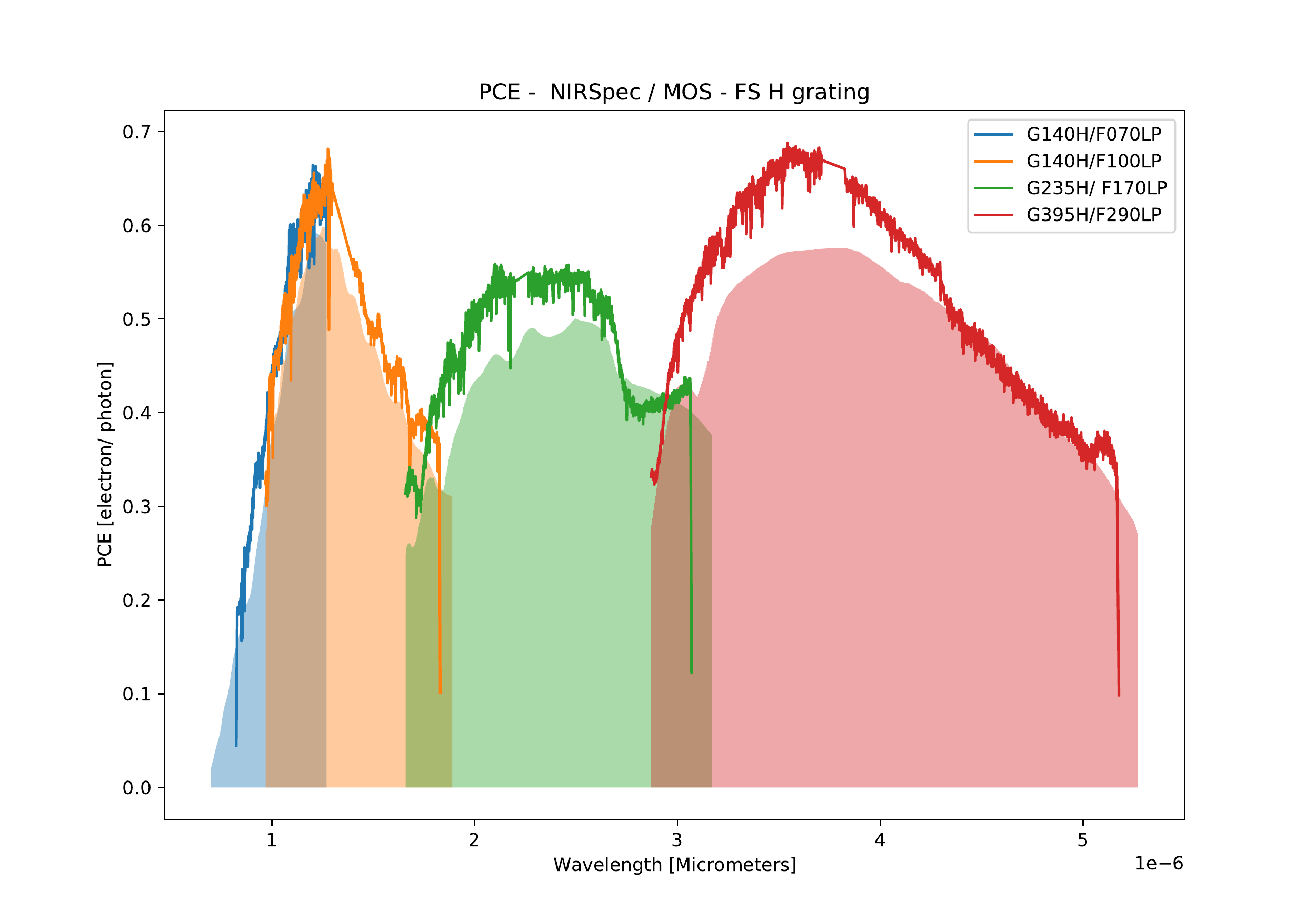}
\end{tabular}
\end{center}
\caption[pce_grating] { \label{fig:gpce} The measured PCE (coloured
  lines) as a function of wavelength of the medium and high
  resolution configurations, for the FS/MOS mode, compared with the values
  predicted by our radiometric model (shaded areas). The model
  excludes light loss occurring in the slit (S1600A1), while the
  measured values are affected by this (for which we have not applied
  any correction)}
\end{figure}

\begin{figure}
\begin{center}
\begin{tabular}{cc} 
  \includegraphics[height=6cm]{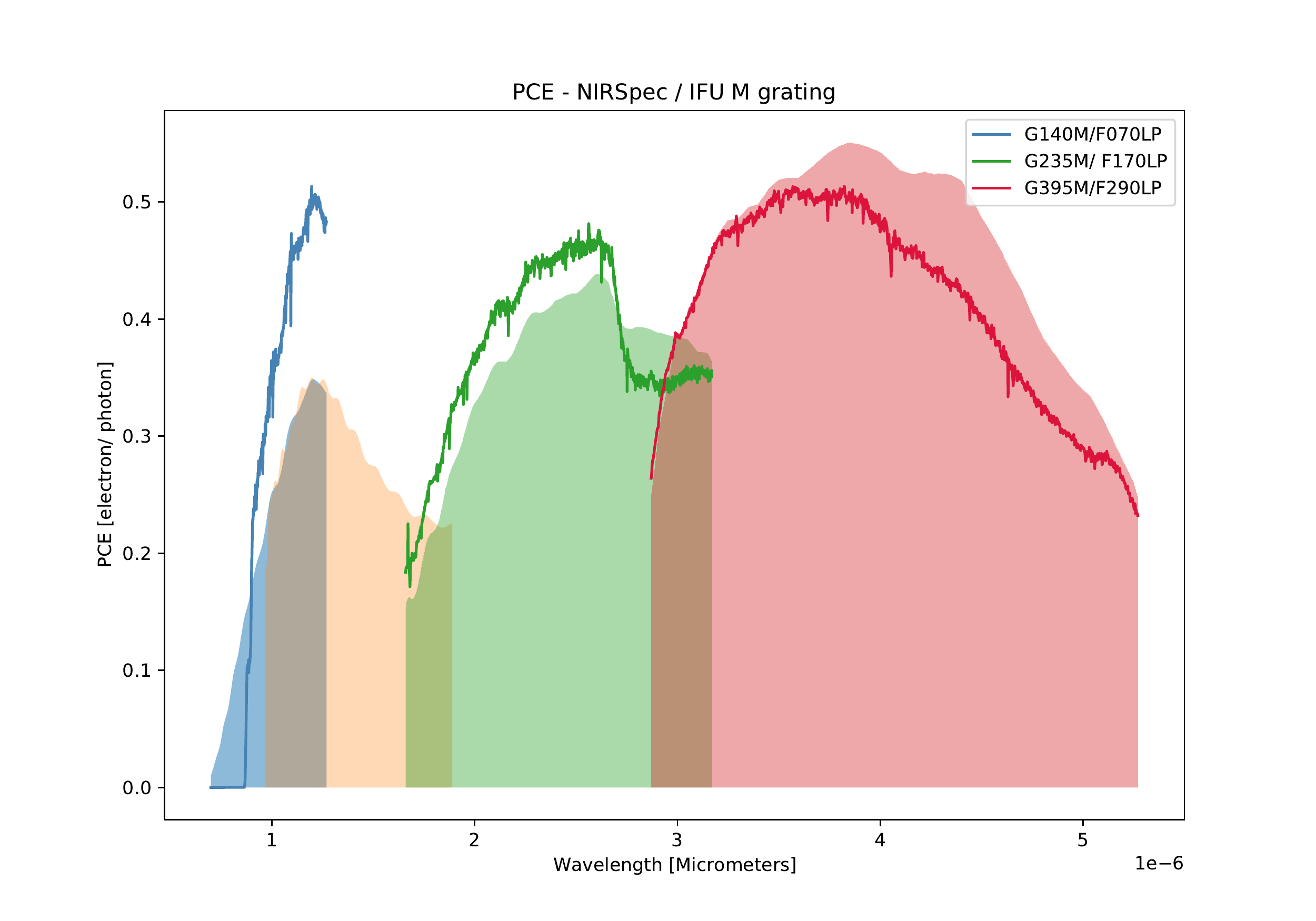}
  \includegraphics[height=6cm]{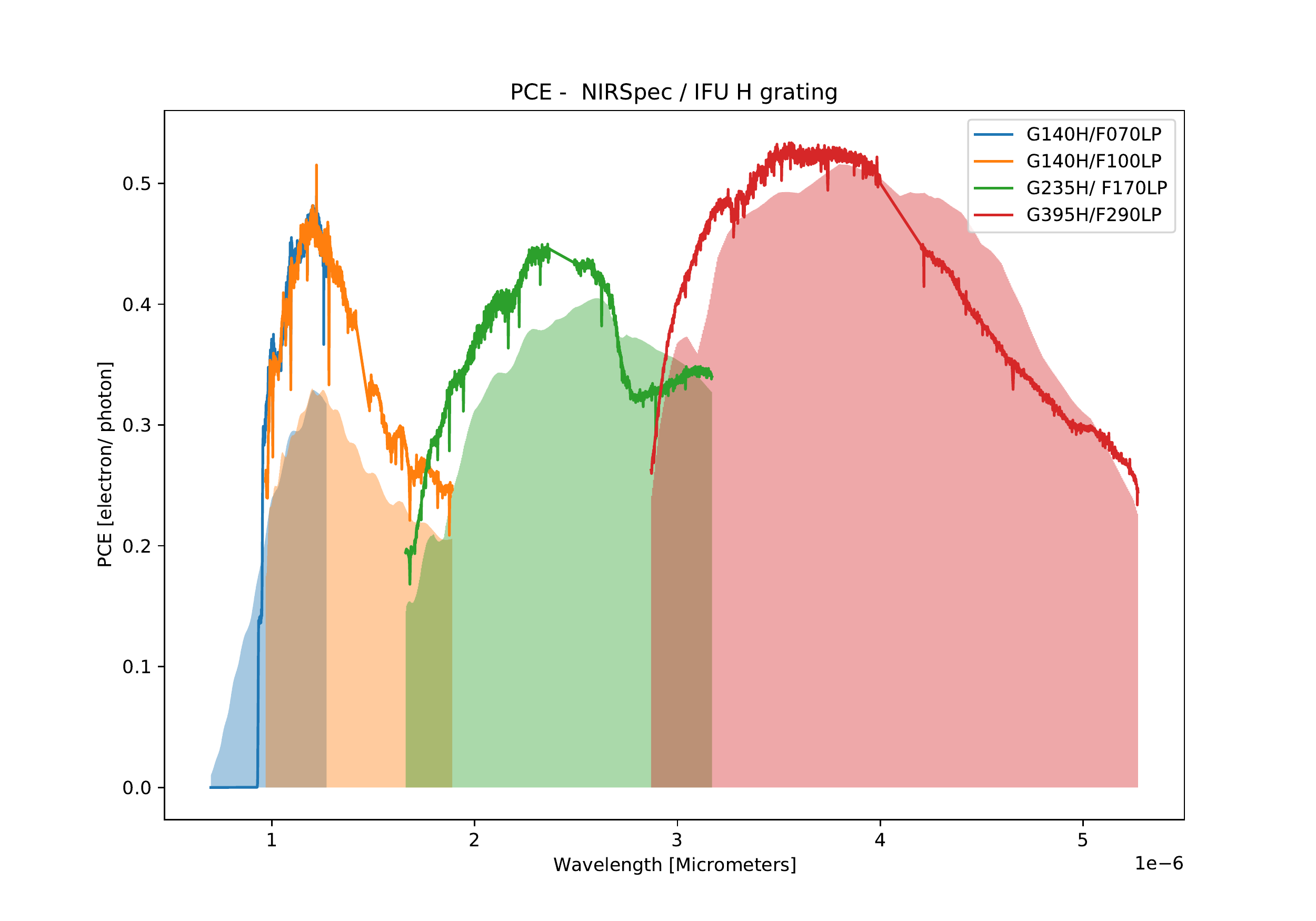}
\end{tabular}
\end{center}
\caption[pce_ifu] { \label{fig:pce_ifu} The measured PCE (coloured
  lines) as a function of wavelength of the medium and high
  resolution configurations, for the IFS mode, compared with the
  values predicted by our radiometric model (shaded areas). The model
  excludes diffraction loss occurring at the IFU slices, while the
  measured values are affected by this (for which we have not applied
  any correction)}
\end{figure}

The PCE for the MSA/FS mode and the IFS mode are shown in
Fig.\,\ref{fig:prismpce} for the low-resolution configuration,
together with predictions from the model of the instrument response
that we developed and used throughout to estimate the sensitivity of
NIRSpec before flight. In the figure, the throughput response of the
different components entering the model are also shown. The model does
not include the light losses due to the finite size of the aperture, which
however do affect the data presented here - in the case of the
square-aperture these are expected to be relatively small.
The PCE values, in MSA/FS mode, for the medium and high resolution configurations are
shown in Fig.\,\ref{fig:gpce} - also compared to the model. For the IFU, the PCE values for the grating configurations available are shown in Fig.\,\ref{fig:pce_ifu}.

Although not presented here, we note that for the configurations for
which observations of more than one standard star was available
we obtained results completely consistent with each other.

In general the efficiency of NIRSpec meets or exceeds the predictions
of our pre-flight model. Significantly higher than predicted
performances are achieved in particular for the high-resolution
configurations, for the MOS/FS mode, and for all configurations below
$\sim2.6~\mu$m, for the IFS mode. We notice a dip in efficiency with
regard to predictions above $\sim 4~\mu$m -- in particular for the IFS
mode, where we see 10--20\% less throughput than predicted. In the
case of the FS/MOS, this difference is (at least partly) explained by
the effect of the slit losses and the fact that they become more
significant at these wavelength, due to the geometrical truncation of
the larger point-spread function (PSF) - see next section. In the case
of IFS mode, the differences between data and predictions are likely due,
partly, to diffraction losses at the level of the IFU 100 mas-slice,
which affect the observations (prevalently in the red) but are not
included in the model and, partly, to slight inaccurate assumptions in
the gold-reflectivity as a function of wavelength, at cryogenic
temperatures (recall that the IFU adds 8 gold-coated reflections to
the light-path).

\subsection{Path-losses}
\label{sec:losses}

Ultimately, in the case of NIRSpec, the total amount of light
registered when observing an object is also affected by the light
losses occurring in the slit device employed. These depend on a
variety of factors: the nature and shape of the target, the slit
employed, the positioning of the object within the aperture, and the
optical quality of the image formed within the slit. For simplicity we
will limit our discussion here to the case of a centered point
source. In this case, when employing the relatively wide square
aperture (S1600A1), the losses are small, but for the narrower slits
and the micro-shutters these are substantial. In the IFS mode, where the
flux of the scientific target in the image plane can be recovered by
summing over multiple slices, the path losses are limited to their
diffraction component and can be in excess of 10\% at the longest
wavelengths.

Before flight, we computed the expected path losses for a point
source, in the various apertures, using a Fourier-optics approach
based on the available measured wavefront error maps of all surfaces
in the light path[\citenum{pfa+2022}], including predictions for the
telescope. Indeed, because the major source of losses is due to the
truncation of the PSF in the silt plane by the geometry of the
aperture, the quality of the image is of fundamental importance.

\begin{figure}
\begin{center}
\begin{tabular}{c} 
  \includegraphics[height=6cm]{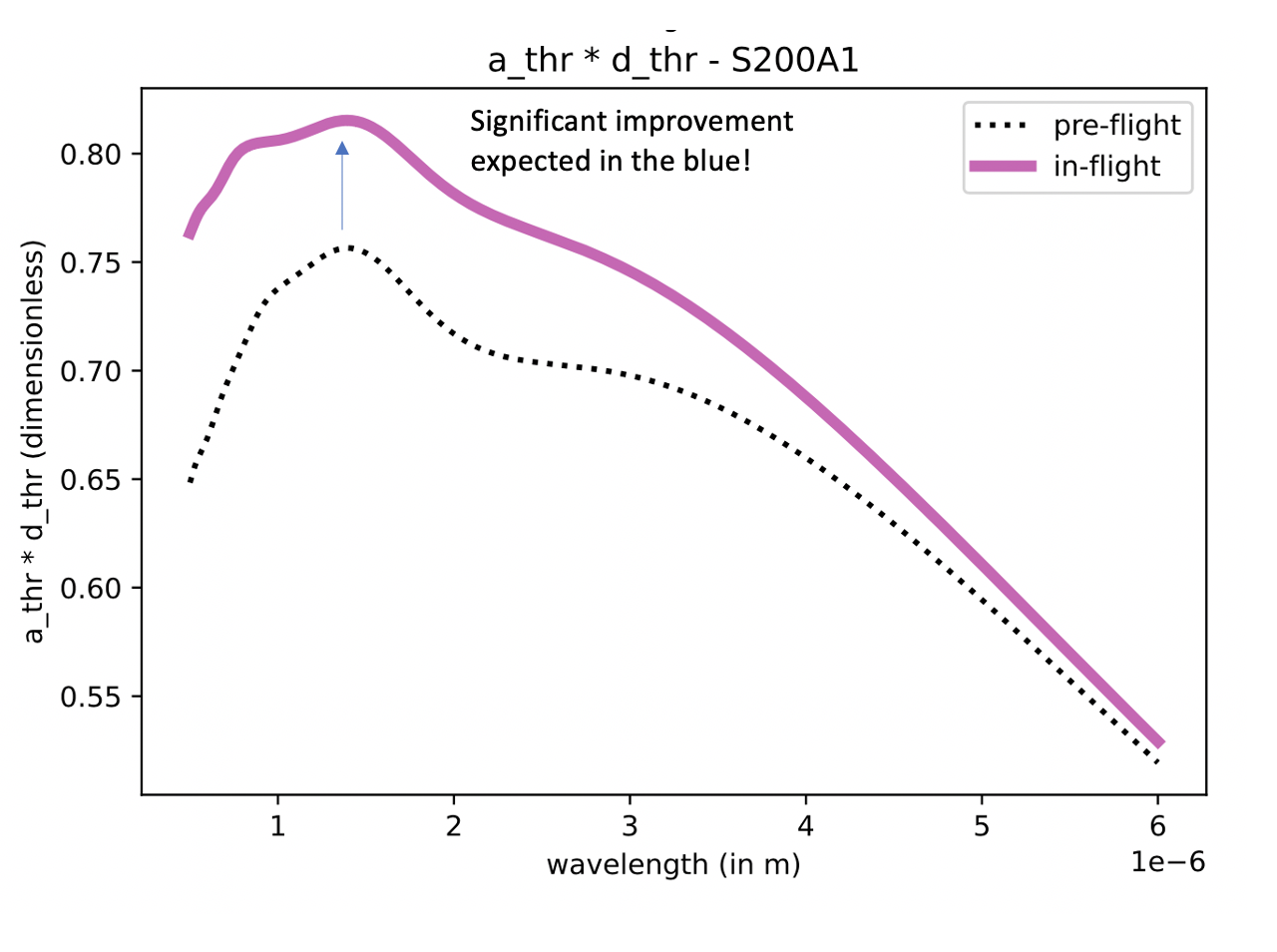}
   \includegraphics[height=6.5cm]{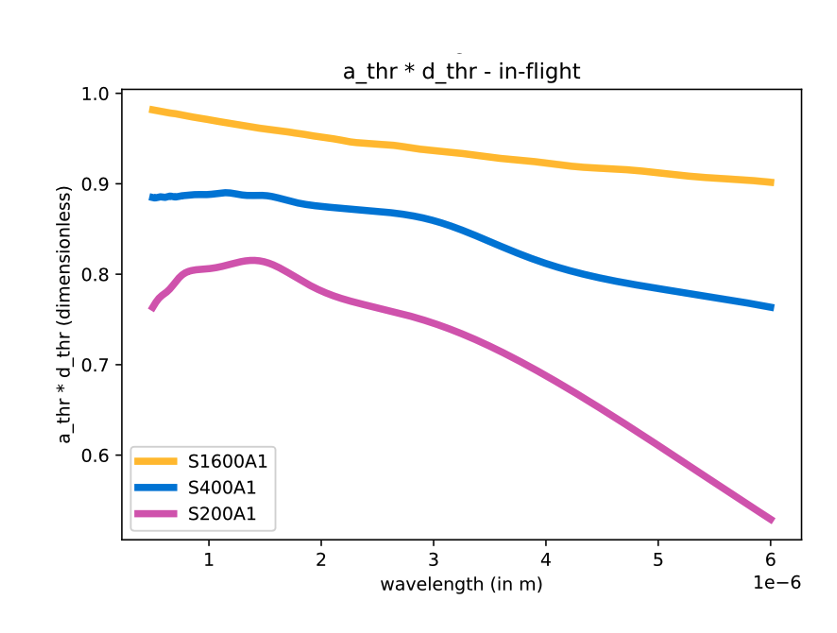}
\end{tabular}
\end{center}
\caption[transm] { \label{fig:stransm} Left-panel: prediction of slit
  transmission pre-flight and in-flight after having updated the
  wavefront-error map of the telescope used in the
  computation. Right-panel: estimated in-flight slit transmission for
  the available slit sizes}
\end{figure} 

Fig.\,\ref{fig:stransm} shows the predicted transmission coefficient
as a function of wavelength for slit S200A1/A2, for the pre-flight
computation and the updated values using a representative map of the
wavefront error obtained during commissioning -- see
[\citenum{otespie2022}]. Since the PSF achieved in flight is of such a
high quality, in particular  with a higher Strehl ratio, the transmission from our
narrowest slit is predicted to improve significantly up to $\sim
4$~$\mu$m, where the truncation of the PSF core starts to be the
dominant factor in the loss of flux.

\begin{figure}
\begin{center}
\begin{tabular}{c} 
  \includegraphics[height=6cm]{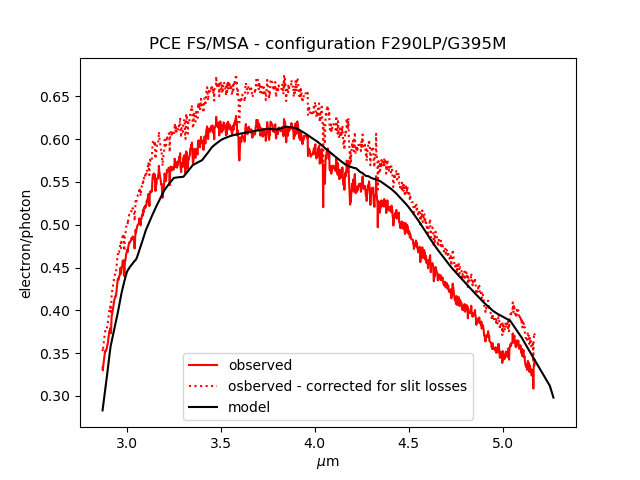}
  \includegraphics[height=6cm]{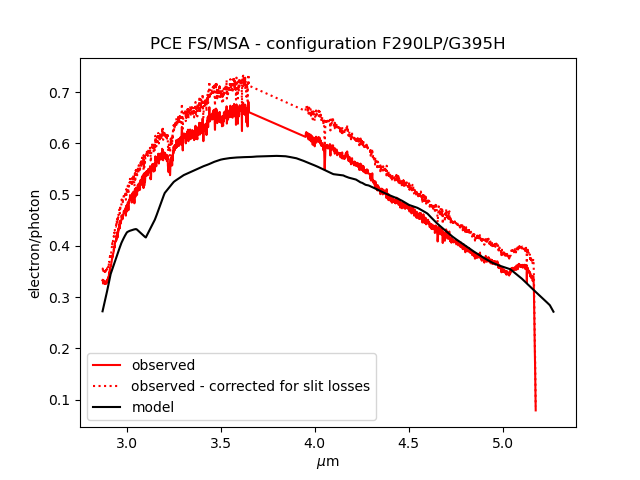}
\end{tabular}
\end{center}
\caption[transm] {\label{fig:corr_pce_g395m} PCE values for Band III,
   through slit S1600A1 as directly observed and after correction by the
  estimated slit transmission coefficients. The pre-flight model-based estimate
  is also shown.}
\end{figure}

The computed transmission coefficients for all the fixed slits as a
function of wavelength based on the in-flight PSF are also shown in
Fig.\,\ref{fig:stransm} - right panel; the transmission coefficients
shown here include both the geometrical losses due to the finite size
of the aperture, the dominant term, and the diffraction losses in the
collimator pupil, contributing by a few percent[\citenum{pfa+2022}].
While the relative improvement of the transmission, with respect to
the pre-flight prediction, is most noticeable in the blue, the
absolute level of losses are higher on the red-side of NIRSpec
wavelength range where the PSF full-width half-maximum is larger:
indeed, this is partly the reason for an apparent drop in performance
in the red when comparing the observed PCE with the model predictions
(which does not include path losses), for the FS/MOS mode.  Using the
transmission coefficients for slit S1600A1, to correct for this effect
for the case of the F290LP filter, both M and H gratings, brings the
PCE values right in line with the predictions also for these
configurations, as shown in Fig.\,\ref{fig:corr_pce_g395m}.

We have currently not yet updated our estimates of path losses in the
case of the IFS mode.

\section{DISCUSSION AND CONCLUSION}

For many years during the development of NIRSpec, we had to rely on a
model of the instrument PCE, anchored to measurements obtained during
the instrument level testing that took place in 2013 - when the
instrument was not yet in its final flight configurations.  For the
MOS/FS mode, the anticipated PCE, excluding slit losses, was very
high, reaching in excess of 50\% for the prism, and at peak blaze in
all six gratings [\citenum{pfa+2022}]. For the IFS the PCE performance
was predicted to be approximately 50\% to 10\%
lower, going from the blue- to the red-end of NIRSpec operating
wavelength range[\citenum{bal+2022}], since the IFS introduces eight
further reflections.

The observations performed during commissioning demonstrates that
NIRSpec is equipped with extraordinarily efficient optics and
detectors. In the MOS/FS mode, the observed PCE overall meets or
exceeds the predictions; for the IFS mode, the performance in the blue
are significantly higher than anticipated ($+$30\% below 2~$\mu$m) and
lower than expected above 4~$\mu$m (up to $-$20\%). In this case, the reasons for
the discrepancy between the model and the measurements could be a
combination of unaccounted for diffraction losses at the level of
the IFU 100 mas-slices and inaccurate assumptions in terms of the
gold-reflectivity as a function of wavelength, at cryogenic
temperatures.

Beside the optical efficiency, the other fundamental parameter
driving the photometric sensitivity of an instrument such as NIRSpec
is the noise performance of the detector. NIRSpec is equipped with
extremely low-noise Teledyne H2RG arrays and, as illustrated by
S. Birkmann et al.[\citenum{st_spie+2022}] at this conference, the
behaviour of the two sensors is excellent, with noise figures that
meet the anticipated in-flight performances. Hence, NIRSpec achieves
the sensitivity projected before launch by Ferruit et
al. [\citenum{fjg+2022}], for the FS and MOS mode, and B{\"o}ker et
al. [\citenum{bal+2022}], for the IFS mode.

JWST/NIRSpec is confirmed to be the most sensitive near-IR
spectrograph currently available for astronomical studies ready to
deliver exciting new observations.

 

\bibliography{nirspec-pce} 
\bibliographystyle{spiebib} 

\end{document}